\documentclass[abstract=on,notitlepage,superscriptaddress,nofootinbib,aps,showpac,prd]{revtex4-2}
\usepackage{amsfonts}
\usepackage{amsmath}
\usepackage{amssymb}
\usepackage{enumerate}
\usepackage{graphicx}
\usepackage{color}
\usepackage{epstopdf}
\usepackage{float}
\usepackage{dcolumn}
\usepackage{hyperref}
\usepackage{amsthm}
\usepackage{comment}

\usepackage{multirow}%BNS table

\begin{document}

%\preprint{}

\title{Gravitational wave interactions with a viscous fluid: Core collapse supernova, binary neutron star merger, and accretion around a black hole merger}



\author{Nigel T. Bishop}\email[]{n.bishop@ru.ac.za}
\affiliation{Department of Mathematics, Rhodes University, Makhanda 6140, South Africa}
\affiliation{National Institute for Theoretical and Computational Sciences (NITheCS), South Africa}

\author{Vishnu Kakkat}\email[]{vishnu.kakkat@nithecs.ac.za}
\affiliation{Department of Mathematical Sciences, University of South Africa, P.O. Box 392, Unisa 0003, Pretoria, South Africa}
\affiliation{National Institute for Theoretical and Computational Sciences (NITheCS), South Africa}
\affiliation{Institutionen för teknikvetenskap och matematik, Luleå tekniska universitet, 971 87 Luleå, Sweden}

\author{Monos Naidoo}\email[]{monos.naidoo@ru.ac.za}
\affiliation{Department of Mathematics, Rhodes University, Makhanda 6140, South Africa}
\affiliation{National Institute for Theoretical and Computational Sciences (NITheCS), South Africa}

%\thanks{}
\begin{abstract}
The interaction of gravitational waves (GWs) with matter is normally treated as being insignificant. However, recent work has shown that the interaction with a viscous fluid may be astrophysically important when the distance between the matter and GW source is somewhat smaller than the GW wavelength. Previous work has mainly considered perturbations on a Minkowski background, and here these results are extended to the case that the background is a general, non-vacuum, static, spherically symmetric spacetime. Expressions are obtained for GW damping and the consequent heating of the fluid, and implemented in computer code. The results are applied to astrophysical scenarios: Core collapse supernovae, the post-merger signal from a binary neutron star merger, and matter accreting at a binary black hole merger. It is found that, compared to the Minkowski case, the damping and heating effects increase, in some cases by several orders of magnitude. It is possible for a GW signal to be completely damped, and for the heating to be such that a gamma-ray burst occurs.

	\end{abstract}
	
%	\keywords{Gravitational waves \and viscosity \and Bondi-Sachs \and Matter shell \and Linearized perturbation theory}%Use showkeys class option if keyword
	\maketitle
	
	%\tableofcontents
	
	\section{Introduction}
	\label{intro}
That gravitational waves (GWs) can transfer energy to matter has been known for many years, since Feynman's presentation of the ``sticky bead'' thought experiment in 1957. Subsequently, in 1966 Hawking~\cite{Hawking1966} considered GWs propagating in various cosmological models, from which it follows that GWs are damped
\begin{equation}
H(r+\Delta)=H(r)\exp\left(-\frac{8\pi G \eta\Delta}{c^3}\right)\,,
\label{e-Haw}
\end{equation}
where $H$ is the magnitude of the GWs (in the $TT$ gauge), $\Delta$ is the distance propagated through a fluid with shear viscosity $\eta$. In SI units, $G/c^3={\mathcal O}(10^{-36})$ and there are no known cosmological or astrophysical scenarios in which $\eta\Delta$ even approaches $10^{36}$. Thus, although it is understood that GWs do interact with matter, it has become standard practice to regard the effect as being insignificant.

Previous work re-investigated this issue in spherical coordinates: using the Bondi-Sachs metric~\cite{Bondi62,Sachs62} we considered small perturbations about a Minkowski background~\cite{Bishop:2022kzq}. Eq.~\eqref{e-Haw} was recovered when the matter is far from the source, but when the distance $r$ between the matter and the GW source satisfies $r\lesssim\lambda$ (where $\lambda$ is the GW wavelength) then the damping effect can be astrophysically significant. Further, if the GWs lose energy through damping then the matter must be heated and this can also be significant~\cite{bishop2024heating, Bishop2024Heating2}. It was found that either of, or both, the damping or heating effects can be significant for: Core collapse supernovae (CCSNe), the post-merger signal from a binary neutron star (BNS) merger, matter accreting at a binary black hole (BBH) merger, and primordial GWs emitted during the inflationary era.

The simplest background to use is Minkowski, but clearly this background does not allow for the GW sources having strong gravitational fields. Ref.~\cite{Bishop2024Schwarzschild} extended previous results to the case that the background is Schwarzschild. This was not a straightforward extension since it gives a system of ODEs that has to be integrated numerically (whereas for a Minkowski background, everything can be done analytically). In this paper, previous results are extended to the case that the background is a general, non-vacuum, static, spherically symmetric spacetime, and computer code is produced that calculates the GW damping and heating effects. Further, the focus of this paper is on astrophysical applications: Core collapse supernovae (CCSNe), the post-merger signal from a binary neutron star (BNS) merger, and matter accreting at a binary black hole (BBH) merger. It is found that the change in background increases the damping and heating effects, in some cases by several orders of magnitude. This emphasizes the importance of including GW damping/heating in astrophysical models where $r\lesssim \lambda$.

Sec.~\ref{s-prev} summarizes the previous work on which this paper is based, then Sec.~\ref{s-model} presents the model and the equations for the various background and perturbed quantities. The results of coding accuracy tests are presented in Sec.~\ref{s-codetest}. The application of the code to some astrophysical scenarios is the subject of Sec.~\ref{s-astrophys}. The paper ends with a Conclusion~\ref{s-conclusion}. Some lengthier formulas are given in Appendix~\ref{a-SI}, and the computer files (available as online supplementary material) are described in Appendix~\ref{a-MapleOctave}.
The theory parts of this paper use geometric units with the gravitational constant $G$ and the speed of light $c$ set to unity, but  SI units are used in the astrophysical sections (\ref{s-GWdh} and \ref{s-astrophys}).
	
	%%%%%%%%%%%%%%%%%%%%%%%%%%%%%%%%%%%%%%%%%%%%%%%%%%%%%%%%%%%%%
	\section{Previous work}
	\label{s-prev}
	
	We use the formalism developed previously~\cite{Bishop:2019eff} on GWs propagating through matter shells. The coordinates are $(x^0,x^1,x^2,x^3)=(u,r,x^A)$ where $x^A$ are angular coordinates and $A=2,3$. The  metric is in Bondi-Sachs form~\cite{Bondi62,Sachs62}
	\begin{align}\label{eq:bmet}
		ds^2  = & -\left(e^{2\beta}\left(1 + \frac{W}{r}\right)
		- r^2h_{AB}U^AU^B\right)du^2
		- 2e^{2\beta}dudr \nonumber \\
		& - 2r^2 h_{AB}U^Bdudx^A
		+  r^2h_{AB}dx^Adx^B\,,
	\end{align}
where $h^{AB}h_{BC}=\delta^A_C$, and the condition that $r$ is a surface area coordinate implies $\det(h_{AB})=\det(q_{AB})$ where $q_{AB}$ is a unit sphere metric (e.g. $d\theta^2+\sin^2\theta d\phi^2$). We represent $q_{AB}$ by a complex dyad (e.g. $q^A=(1,i/\sin\theta)$) and introduce the complex differential angular operators $\eth,\bar{\eth}$~\cite{Newman-Penrose-1966}, with the operators defined with respect to the unit sphere as detailed in~\cite{Bishop2016a,Gomez97}. Then $h_{AB}$ is represented by the complex quantity $J=q^Aq^Bh_{AB}/2$ (with $J=0$ characterizing spherical symmetry), and we also introduce the complex quantity $U=U^Aq_A$.

We now make the ansatz of small perturbations about a general, spherically symmetric, static background, explicitly
	\begin{align}
		\beta=&\beta^{[B]}(r)+\Re(\beta^{[2,2]}(r)e^{i\nu u}){}_0Z_{2,2}\,,\;\;
		U=\Re(U^{[2,2]}(r)e^{i\nu u}){}_1Z_{2,2}\,,\nonumber \\
		W&=W^{[B]}(r)+\Re(W^{[2,2]}(r)e^{i\nu u}){}_0Z_{2,2}\,,\;\;
		J=\Re(J^{[2,2]}(r)e^{i\nu u}){}_2Z_{2,2}\,,
		\label{e-ansatz}
	\end{align}
with background quantities denoted by the superscript ${}^{[B]}$. Previous work has studied the cases that the background is Minkowski~\cite{Bishop:2022kzq} for which $\beta^{[B]}=W^{[B]}=0$, and Schwarzschild~\cite{Bishop2024Schwarzschild} where $\beta^{[B]}=0,W^{[B]}=-2M$. The perturbations oscillate in time with frequency $\nu/(2\pi)$. The quantities ${}_s Z_{\ell,m}$ are spin-weighted spherical harmonic basis functions related to the usual ${}_s Y_{\ell,m}$ as specified in~\cite{Bishop-2005b,Bishop2016a}. They have the property that ${}_0 Z_{\ell,m}$ are real, enabling the description of the metric quantities $\beta,W$ (which are real) without mode-mixing; however, for $s\ne 0$ ${}_s Z_{2,2}$ is, in general, complex. A general solution may be constructed by summing over the $(\ell,m)$ modes, but that is not needed here, since we are considering a source that is continuously emitting GWs at constant frequency dominated by the $\ell=2$ (quadrupolar) components.

\section{Model for perturbations on a spherically symmetric background with matter}
\label{s-model}

\subsection{Background solution}
Firstly, the unperturbed, or background, model needs to be specified. The background is static and spherically symmetric with an inner core of mass $M_c$ and radius $r_c$. The region $r_c<r<r_t$ contains fluid which is static and specified by its density $\rho^{[B]}$ and pressure $p^{[B]}$ only. The region $r_t<r$ is vacuum and is described by the Schwarzschild geometry with mass $M_t$ being the total mass in $r<r_t$, i.e. with $W^{[B]}=-2M_t$.

The next step is to determine the background geometry in $r_c<r<r_t$. The density $\rho^{[B]}$ is regarded as given, and the Einstein equations are used to determine $\beta^{[B]},W^{[B]},p^{[B]}$. From the background metric, it follows that the background 4-velocity is
\begin{equation}
V^{[B]}_a=\left(-\exp(\beta^{[B]})\left(1+\frac{W^{[B]}}{r}
\right)^{\frac12},-\exp(\beta^{[B]})\left(1+\frac{W^{[B]}}{r}
\right)^{-\frac12},0,0\right)\,.
\end{equation}
The stress-energy tensor of a perfect fluid is
\begin{equation}
T_{ab}=(\rho+p)V_aV_b+pg_{ab}\,,
\end{equation}
and the Einstein equations are
\begin{equation}
E_{ab}=R_{ab}-8\pi T_{ab}+4\pi g_{ab}T=0\mbox{  where  }T=T_{ab}g^{ab}\,.
\end{equation}

Then the Einstein equations $E_{11},g^{AB}E_{AB}$, together with the matter conservation condition $\nabla_cT_{1b}g^{bc}=0$, give
\begin{align}
\partial_r \beta^{[B]} &=2\pi r^2 \frac{(\rho^{[B]}+p^{[B]})\exp(2\beta^{[B]})}{r+W^{[B]}}\nonumber \\
\partial_r W^{[B]}&=\exp(2\beta^{[B]})-1-4\pi r^2\exp(2\beta^{[B]})(\rho^{[B]}-p^{[B]})\nonumber \\
\partial_r p^{[B]}&=-\left(r\partial_rW^{[B]}+2\partial_r\beta^{[B]}(r^2+rW^{[B]})-W^{[B]}\right)\frac{\rho^{[B]}+p^{[B]}}{2r(r+W^{[B]})}\,.
\label{e-sys_back}
\end{align}

Eqs.\eqref{e-sys_back} constitute a system of 3 first-order ODEs, and are solved numerically in the range $r_c<r<r_t$ using the exterior solution to provide boundary data, explicitly $\beta^{[B]}=p^{[B]}=0,W=-2M_t$ at $r=r_t$. The accuracy of the numerical procedure is checked by comparing results to those of an analytic solution, namely the interior Schwarzschild solution, see Sec.~\ref{s-codetest}.

\subsection{Perturbed solution}

In the exterior region $r>r_t$ the background geometry is Schwarzschild and the perturbed solution obtained in~\cite{Bishop2024Schwarzschild} is used. The solution is obtained numerically, and thus requires that the gauge be fixed: we use the ``Bondi'' gauge in which the metric is explicitly asymptotically flat and all metric terms fall off at least as fast as $1/r$. In particular $\beta^{[2,2]}(r)=0$. This solution is used to find
\begin{equation}
J^{[2,2]}(r),\,\partial_r J^{[2,2]}(r), U^{[2,2]}(r),\,\partial_r U^{[2,2]}(r),\mbox{ and }W^{[2,2]}(r)
\label{e-pertvars}
\end{equation}
at $r=r_t$, and this data will be used as boundary conditions for constructing a numerical solution in $r_c<r<r_t$. In this region, the Einstein equations are
\begin{align}
E_{11}:&\rightarrow \partial_r \beta^{[2,2]}=0\,,\nonumber\\
q^AE_{1A}:&\rightarrow (2r^2\partial_r\beta^{[B]}-4r)\partial_rU^{[2,2]}
-r^2\partial^2_rU^{[2,2]}+4\exp(2\beta^{[B]})\partial_rJ^{[2,2]}=0\,,\nonumber\\
q^Aq^BE_{AB}:&\rightarrow \partial_r(r^2U^{[2,2]})-\partial_r(rW^{[B]}\partial_rJ^{[2,2]})-\partial_r(r^2\partial_rJ^{[2,2]})+2i\nu r\partial_r(rJ^{[2,2]})=0\,,\nonumber\\
h^{AB}E_{AB}:&\rightarrow \partial_rW^{[2,2]}+12rU^{[2,2]}+3r^2\partial_rU^{[2,2]}
-12\exp(2\beta^{[B]})J^{[2,2]}=0\,,
\label{e-pertsys}
\end{align}
and can be expressed as a system of five ODEs in the variables listed in Eq.~\eqref{e-pertvars}. The system is solved numerically. The accuracy of the numerical procedure is checked by setting the shell density to be very small and comparing results to those previously obtained on a Schwarzschild background, see Sec.~\ref{s-codetest}. It should be noted that the matter code integrates numerically a system of five ODEs, whereas in the Schwarzschild code the equations for $q^AE_{1A}$ and $q^Aq^BE_{AB}$ reduce to a single second order master ODE, which is solved numerically, and then the metric variables reduce to expressions involving only quadrature. Since a numerical quadrature is more accurate than a numerical ODE solution, it may be expected that the matter code will be a little less accurate than the Schwarzschild code.

\subsection{Fluid shear}
As was done for the metric, the velocity is decomposed into background plus perturbations, $V_a=V_a^{[B]}+v_a$ where
\begin{equation}
v_0=\Re(v_0^{[2,2]}(r)e^{i\nu u}){}_0Z_{2,2}\,,\;
v_1=\Re(v_1^{[2,2]}(r)e^{i\nu u}){}_0Z_{2,2}\,,\;
q^Av_A=\Re(V_{ang}^{[2,2]}(r)e^{i\nu u}){}_1Z_{2,2}\,.
\end{equation}
The condition $V_aV_bg^{ab}=-1$ simplifies to give an explicit expression for $v^{[2,2]}_0$
\begin{equation}
v^{[2,2]}_0=-\frac{\exp(\beta^{[B]})W^{[2,2]}}{2\sqrt{r^2+rW^{[B]}}}\,.
\label{e-v0}
\end{equation}
Then the matter conservation condition $q^A\nabla_cT_{Ab}g^{bc}=0$ leads to an explicit expression for $V^{[2,2]}_{ang}$
\begin{equation}
V^{[2,2]}_{ang}=\frac{i\exp(\beta^{[B]})W^{[2,2]}}{2\nu \sqrt{r^2+rW^{[B]}}}\,.
\label{e-Vang}
\end{equation}

The condition for $v^{[2,2]}_1$ is more complicated. The matter conservation conditions $\nabla_cT_{0b}g^{bc}=0$, $\nabla_cT_{1b}g^{bc}=0$ lead to a pair of equations involving $\rho^{[2,2]}$, $v^{[2,2]}$ and $\partial_r v^{[2,2]}$. Then, eliminating $\rho^{[2,2]}$ from the system gives
\begin{equation}
(c_{1,\rho}c_{0,v1}-c_{0,\rho}c_{1,v1})\partial_r v^{[2,2]}_1
=c_{0,\rho}c_{1,0}-c_{1,\rho}c_{0,0} +(c_{0,\rho}c_{1,v}-c_{1,\rho}c_{0,v})v^{[2,2]}_1\,,
\label{e-dv1}
\end{equation}
where some of the coefficients involve lengthy expressions and are given in the computer code, see Appendix~\ref{a-MapleOctave}.
In the case that the background is Minkowski, the coefficient of $\partial_r v^{[2,2]}_1$ vanishes and Eq.~\eqref{e-dv1} reduces to an explicit expression for $v^{[2,2]}_1$. In the general case, Eq.~\eqref{e-dv1} is an ODE in $v^{[2,2]}_1$ and is solved numerically using the Minkowski value as boundary condition.

Once the velocity field is determined, the fluid shear $\sigma_{ab}$ is evaluated using the definitions
	\begin{align}
		\Theta&=g^{ab}\nabla_a V_b\,,\;\; P_{ab}=g_{ab}+V_a V_b\,,\nonumber \\
		\sigma_{ab}&=\frac{(P_{ac}\nabla_d V_b +P_{bc}\nabla_dV_a)g^{cd}}{2}-\frac{P_{ab}\Theta}{3}\,,
	\end{align}
from which the shear scalar $\sigma^2=\sigma_{ab}\sigma_{cd}g^{ac}g^{bd}$ is calculated using computer algebra, see Appendix~\ref{a-MapleOctave}. Then the rate of energy transfer per unit volume to the fluid is $2\eta\sigma^2$ where $\eta$ is the shear viscosity~\cite{Baumgarte2010a}. The time dependence of each term in a component of $\sigma_{ab}$ is $\cos(\nu u)$ or $\sin(\nu u)$, so that $\sigma^2$ involves terms $\cos^2(\nu u)$, $\sin^2(\nu u)$ and $\cos(\nu u)\sin(\nu u)$.  Then $\sigma^2$ is greatly simplified on taking the time-average, since this reduces $\cos^2(\nu u)$ and $\sin^2(\nu u)$ to $0.5$, and $\cos(\nu u)\sin(\nu u)$ to $0$; note that a consequence of time-averaging is that the results to be obtained are applicable only to processes with a time-scale longer than the wave period.

\subsection{GW damping and heating}
\label{s-GWdh}
We integrate the energy transfer over the sphere to obtain a formula for the rate of energy transfer from GWs to the fluid for a shell of radius $\delta r$ (Note that the integration over the sphere is almost trivial because of the orthonormality of the angular basis functions):
\begin{equation}
\dot{E}=\frac{16\pi\eta G}{c^3}\delta r f_E\,,
\label{e-fE}
\end{equation}
where the quantity $f_E$ is evaluated numerically. In the case that the background is Minkowski, $f_E$ has a simple form denoted by $f_{E,M}$
\begin{equation}
f_{E,M}=1+\frac{\lambda^2}{2\pi^2r^2}+\frac{9\lambda^4}{16\pi^4r^4}+\frac{45\lambda^6}{64\pi^6r^6}+\frac{315\lambda^8}{256\pi^8r^8}\,.
\label{e-fEM}
\end{equation}
Following~\cite{Bishop:2022kzq}, the GW damping effect is expressed as
\begin{equation}
H_o=H_i\exp\left(-\frac{8\pi G\eta}{c^3}\int_{r_i}^{r_o} f_E(r) dr\right)\,,
\label{e-Ho}
\end{equation}
where $\eta$ is the dynamical shear viscosity, and $H_o,H_i$ are the rescaled gravitational wave strains at $r_o,r_i$ respectively; i.e. $r(h_++ih_\times)=H\Re(e^{i\nu u}){}_2Z_{2,2}$.

The heating effect on a Minkowski background is described in~\cite{bishop2024heating, Bishop2024Heating2}. In the case that it is appropriate to average over the spherical shell
\begin{equation}
\Delta T=\frac{2G\eta \Delta E_{GW} f_E}{C\rho c^3r^2}\,,
\label{e-DeltaT}
\end{equation}
where $\Delta E_{GW}$ is the energy transported through the shell in a given time-interval, $C$ is the specific heat of the matter in the shell and $\rho$ is its density. The above formula applies when the thermal diffusivity is high, or when there is no information about the orientation of the GW source. Ref.~\cite{bishop2024heating} evaluated the heating effect when the GW source is a circular binary, obtaining
\begin{equation}
\Delta T=\frac{\pi G \nu^2\eta\Delta E_{GW}}{3\rho c^5 C}\left(D_0 Y_{00}+D_2Y_{20}+D_4Y_{40}\right)\,,
\end{equation} 
where $Y_{\ell m}$ are standard spherical harmonics, and formulas for the $D_a$, in the Minkowski case, are given in~\cite{bishop2024heating}. Here, the $D_a$ are evaluated numerically as follows. Denoting a time-average by $\left<\right>$, we have
\begin{equation}
\frac{\left<\sigma^2\right>r^2}{6\nu^2}=f_{E0} \,{}_0Z_{22}^2+f_{E1} \,{}_1Z_{22}\,{}_{-1}Z_{22}+f_{E2} \,{}_2Z_{22}\,{}_{-2}Z_{22}\,;
\label{e-fE012}
\end{equation}
note that ${}_sZ_{22}$ and ${}_{-s}Z_{22}$ are complex conjugates so this expression is real. Integrating over the sphere and using the orthonormality property of the ${}_sZ_{22}$ leads to $f_E=f_{E0}+f_{E1}+f_{E2}$. The expressions for the $f_{Ea}$ are lengthy, and are given in the computer algebra output and evaluated numerically. Expressions for the $D_a$ are then:
\begin{equation}
D_0=\frac{12c^2(f_{E0}+f_{E1}+f_{E2})}{\sqrt{\pi}r^2\nu^2}\,,
D_1=-\frac{12\sqrt{5}c^2(2f_{E0}+f_{E1}-2f_{E2})}{7\sqrt{\pi}r^2\nu^2}\,,
D_2=\frac{2c^2(6f_{E0}-4f_{E1}+f_{E2})}{7\sqrt{\pi}r^2\nu^2}\,.
\end{equation}

%D0=12*(fE2+fE1+fE0)/sqrt(pi)./rnu.^2;
%D2=-12*sqrt(5)*(2*fE0+fE1-2*fE2)/7/sqrt(pi)./rnu.^2;
%D4=2*(fE2-4*fE1+6*fE0)/7/sqrt(pi)./rnu.^2;

\section{Code testing results}
\label{s-codetest}
\begin{figure}
\includegraphics[scale=0.5]{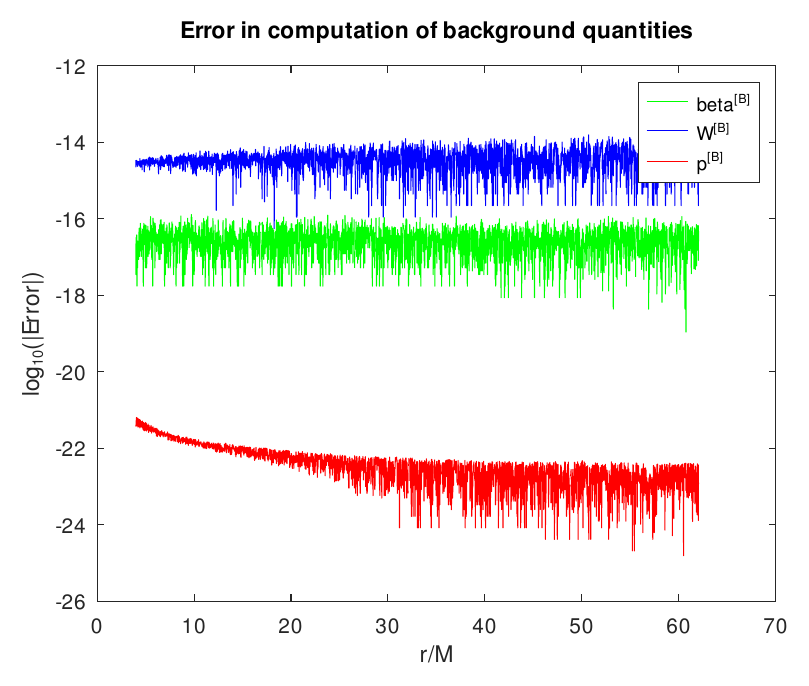}
\caption{Errors in background quantities when the Schwarzschild interior solution applies.}
\label{f-cons}
\end{figure}

The background solution is obtained by integrating the system of ODEs in Eqs.~\eqref{e-sys_back}. In the case that the density $\rho$ is constant, the solution can be compared to the Schwarzschild interior solution~\cite{Schutz85}, and is given in Bondi-Sachs form in Appendix~\ref{a-SI}. The test was performed with $M_t=2.9M_\odot$, $r_t=265.71$km, $r_c=17.13$km, and $\rho^{[B]}=7.33988\times 10^{13}$kg/m${}^3$. The code uses geometric units with $M_t=1$, and in these units $r_t=62.06$, $r_c=4$, and $\rho^{[B]}=10^{-6}$. The values for the exact solution over the range $(r_c,r_t)$ were: $\beta^{[B]}$ between  $-0.0123$ and $0$, $W^{[B]}$ between $-0.0976$ and $-2$, and $p^{[B]}$ between $8.29\times 10^{-9}$ and $0$. Fig.~\ref{f-cons} plots $\log_{10}(|$error$|)$ for $\beta^{[B]},W^{[B]}$ and $p^{[B]}$. In all cases, the errors are less than about $10^{-14}$.

\begin{figure}
\includegraphics[scale=0.5]{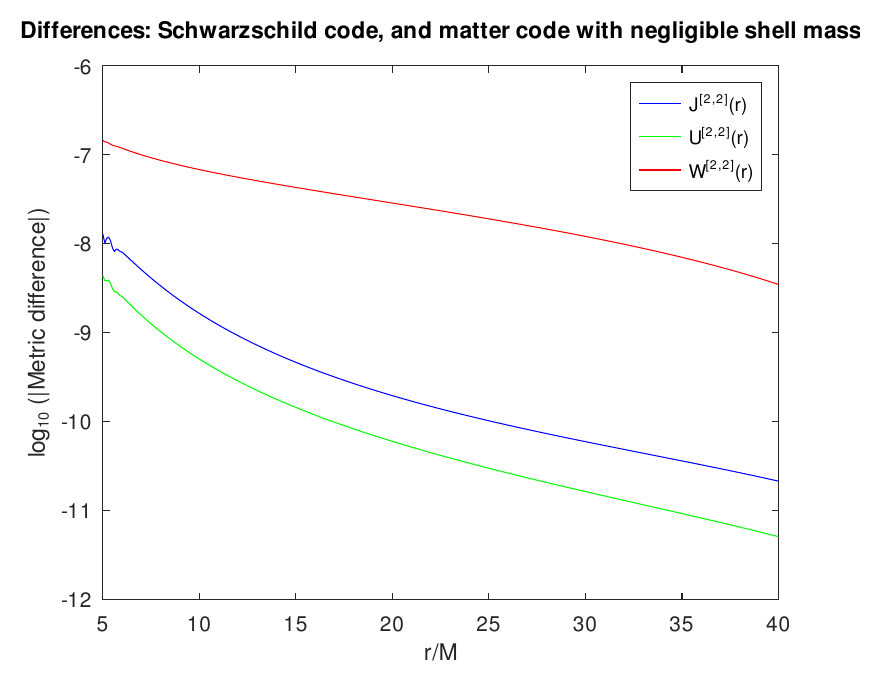}
\includegraphics[scale=0.5]{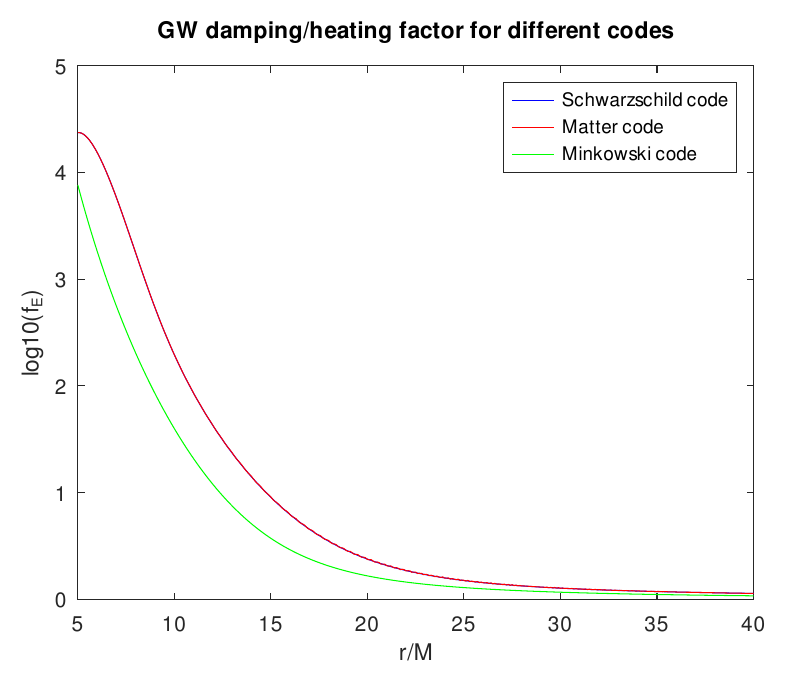}
\caption{Comparisons between the Schwarzschild code and the matter code with negligible density shell. The parameter $\nu=0.1355$, corresponding to a frequency of $72.5$Hz when $M=60M_\odot$. The left panel plots on a $\log_{10}$ scale the differences between the metric quantities indicated for the two codes. The right panel plots $\log_{10}(f_E)$ as calculated by the two codes as well as in the Minkowski background case (Eq.~\eqref{e-fEM}).}
\label{f-Spert}
\end{figure}

Previous work~\cite{Bishop2024Schwarzschild} obtained a numerical solution for the perturbed metric  on a Schwarzschild background, and the current code recovers that case by setting the density of the matter shell to be very small. Fig.~\ref{f-Spert} plots $\log_{10}$ of the absolute value of the difference between the values of $J^{[2,2]},U^{[2,2]}$ and $W^{[2,2]}$ found by the two codes. The integration is from the outer boundary inwards, and it is seen that the differences grow as $r/M$ is decreased, reaching a maximum of order $10^{-7}$. The ODE integration is down to $r/M=3.5$ (which is well away from the horizon at $r/M=2$ where the ODEs become singular) but sufficient for the astrophysical applications to be considered here. We also compare the quantity $f_E$ as calculated by the matter code with negligible density shell and the Schwarzschild code in Fig.~\ref{f-Spert}. It is seen that the two curves are almost on top of each other, and it is notable that for small $r/M$ the value of $f_E$ is significantly larger than in the Minkowski background case (Eq.~\eqref{e-fEM}). In all 3 cases, $f_E\rightarrow 1$ as $r/M\rightarrow\infty$.
 
\section{Astrophysical results}
\label{s-astrophys}

	\subsection{Core-collapse supernovae}
	Core-collapse supernovae (CCSNe) represent one of the most violent astrophysical processes in the Universe. As the iron core of a massive star exceeds the Chandrasekhar mass ($\sim 1.4\,M_{\odot}$), electron degeneracy pressure fails, triggering gravitational collapse \cite{Bethe:1990mw,Janka:2012wk,Jerkstrand:2025bea}. The collapse halts when nuclear densities ($\rho \sim 2-3 \times 10^{17}\,\mathrm{kg\,m^{-3}}$) are reached, leading to the formation of a proto-neutron star (PNS) and an outgoing shock wave. The post-bounce hydrodynamics and anisotropic mass motions produce gravitational waves expected to be detectable by current observatories \cite{Yakunin:2017tus,Andresen:2016pdt}.
	
	The PNS consists of a dense core surrounded by a less dense, hot mantle (or thin shell). The inner core typically has a mass of $M_{c} \sim 0.5$--$0.7\,M_{\odot}$ and radius $r_{c} \sim 10$--$15\,\mathrm{km}$, while the outer mantle (or thin shell) extends to $r_{t} \sim 30$--$50\,\mathrm{km}$ with a mass of $\sim 0.1$--$0.3\,M_{\odot}$ \cite{Nevins:2024dkr, OConnor:2011pxx}. The density falls steeply outside the core, following a power-law or exponential decay.
	
%	A commonly used approximation for the radial density profile of the PNS and its immediate surroundings is:
%	\begin{equation}
%		\rho(r) = \rho_c \left[1 - \left(\frac{r}{R_{\mathrm{core}}}\right)^2 \right]^n,
%		\label{eq:density_profile}
%	\end{equation}
%	where $\rho_c$ is the central density, $R_{\mathrm{core}}$ is the core radius, and $n$ is a structural index typically between 1 and 3 for neutron star cores \cite{Lattimer:2006xb}. Near the outer layers, a steeper exponential tail can also be used:
%	\begin{equation}
%		\rho(r) = \rho_0 \exp\left(-\frac{r-R_{\mathrm{core}}}{H_\rho}\right),
%	\end{equation}
%	where $H_\rho$ is the density scale height, often a few kilometres.
%	
%	\begin{table}[htbp]
%		\centering
%		\begin{tabular}{l|c|c|c}
%			\hline
%			Region & Radius (km) & Mass ($M_{\odot}$) & Density (kg m$^{-3}$) \\
%		\hline
%			Inner core & 10--15 & 0.5--0.7 & $(2$--$3)\times10^{17}$ \\
%			Outer shell (mantle) & 30--50 & 0.1--0.3 & $(1$--$5)\times10^{15}$ \\
%		\hline
%		\end{tabular}
%		\caption{Approximate structural parameters of a proto-neutron star and its thin outer shell.}
%		\label{tab:ccsne_parameters_OLD}
%	\end{table}
	
\begin{table}[t]
\centering
\caption{Parameter values for CCSNe.}
\label{tab:ccsne_parameters}
\begin{tabular}{lc}
\hline
\textbf{Quantity} & \textbf{Typical Range or Value} \\
\hline
Inner core mass & $0.5$--$0.7~M_\odot$ \\
Inner core radius & $10$--$15~\mathrm{km}$ \\
Inner core density & $(2$--$3)\times 10^{17}~\mathrm{kg/m^3}$ \\
Outer shell (mantle) mass & $0.1$--$0.3~M_\odot$ \\
Outer shell radius & $30$--$50~\mathrm{km}$ \\
Outer shell density & $(1$--$5)\times 10^{15}~\mathrm{kg/m^3}$ \\
Shear viscosity $\eta$ & $10^{23}$--$10^{25}~\mathrm{kg\,m^{-1}\,s^{-1}}$ \\
Specific heat $C$ & $6$--$200~\mathrm{J\,kg^{-1}\,K^{-1}}$ \\
GW frequency & $100$--$1000~\mathrm{Hz}$\\
GW energy release $\Delta E_{\rm GW}$ & $10^{36}$--$10^{39}~\mathrm{J}$ \\
\hline
\end{tabular}
\end{table}
	
	The collapse proceeds on a timescale of $\sim 200$--$400\,\mathrm{ms}$ from onset to core bounce \cite{Muller:2020ard}. When nuclear densities are reached, the inner core stiffens, leading to a rebound that launches a shock wave. This \textit{bounce} occurs within $\sim 1\,\mathrm{ms}$, producing a sharp burst of gravitational radiation \cite{Ott:2012jq}.
	
	Following bounce, convection and the standing accretion shock instability (SASI) drive stochastic mass motions and asymmetric neutrino emission, both contributing to longer-duration GW signals lasting up to several hundred milliseconds \cite{Kuroda:2016bjd,Andresen:2018aom}. The frequency spectrum evolves from $\sim 100\,\mathrm{Hz}$ at early post-bounce phases to $\sim 1$--$1.5\,\mathrm{kHz}$ as the PNS contracts and oscillates in $g$- and $f$-modes.
	
	The GW strain amplitude from a CCSN at 10\,kpc is expected to be $h \sim 10^{-22}$--$10^{-21}$, peaking near bounce and modulated by subsequent PNS oscillations \cite{Yakunin:2017tus}. Frequency bands depend strongly on the mass and rotation of the progenitor:
	\begin{itemize}
		\item Bounce signal: $\sim 500$--$1000\,\mathrm{Hz}$, duration $\sim 10\,\mathrm{ms}$.
		\item Convection/SASI: $\sim 100$--$300\,\mathrm{Hz}$, lasting $\sim 300\,\mathrm{ms}$.
		\item PNS $g$-mode oscillations: $\sim 1$--$2\,\mathrm{kHz}$.
	\end{itemize}
	The GWs from CCSNe encode the dynamics of the forming NS. Their frequencies and amplitudes reflect the density profile, rotation, and equation of state of the collapsing core. As current detectors approach sensitivity to Galactic events, CCSNe remain prime targets for multi-messenger astrophysics.

%%%%%%%%%%%%%%%%%%%%%%%%%%%%%%%%%%%%%%%%

%%%%%%%%%%%%%%%%%%%%%%%%%%%%%%%%%%%%%%%%
	
\begin{table}[h]
	\centering
		\caption{The table shows the GW damping factor at the outer boundary and the temperature rise at the inner-boundary for various CCSNe models, using both the matter Minkowski code.s}

	\renewcommand{\arraystretch}{1.4}
	\begin{tabular}{c c l c c c}
		\hline
		$\eta$ & Frequency & Model & $H_{\textrm{damp}}$ (outer) & $\Delta T_{\textrm{max}}$ (inner) & $\Delta T_{\textrm{min}}$ (inner) \\
		\hline
		
		% ====================== eta = 10^23 ======================
		\multirow{4}{*}{$10^{23}$}
		
		& $\nu = 100\,\text{Hz}$ 
		& Matter code 
		& $<10^{-3}$ 
		& $6.03\times 10^{18}$
		& $1.81\times 10^{14}$ \\
		
		& 
		& Minkowski code 
		& $<10^{-3}$ 
		& $2.87\times 10^{18}$
		& $8.61\times 10^{13}$ \\
		
		& $\nu = 1\,\text{kHz}$ 
		& Matter code 
		& $0.02$ 
		& $4.75\times 10^{10}$ 
		& $1.42\times 10^{6}$ \\
		
		& 
		& Minkowski code 
		& $0.48$ 
		& $2.89\times 10^{10}$
		& $8.66\times 10^{5}$ \\
		\hline
		
		% ====================== eta = 10^24 ======================
		\multirow{4}{*}{$10^{24}$}
		
		& $\nu = 100\,\text{Hz}$ 
		& Matter code 
		& $<10^{-3}$ 
		& $6.03\times 10^{19}$
		& $1.81\times 10^{15}$ \\
		
		& 
		& Minkowski code 
		& $<10^{-3}$ 
		& $2.87\times 10^{19}$
		& $8.61\times 10^{14}$ \\
		
		& $\nu = 1\,\text{kHz}$ 
		& Matter code 
		& $<10^{-3}$ 
		& $4.75\times 10^{11}$
		& $1.42\times 10^{7}$ \\
		
		& 
		& Minkowski code 
		& $<10^{-3}$ 
		& $2.89\times 10^{11}$
		& $8.66\times 10^{6}$\\
		\hline
		
		% ====================== eta = 10^25 ======================
		\multirow{4}{*}{$10^{25}$}
		
		& $\nu = 100\,\text{Hz}$ 
		& Matter code 
		& $<10^{-3}$ 
		& $6.03\times 10^{20}$
		& $1.81\times 10^{16}$ \\
		
		& 
		& Minkowski code 
		& $<10^{-3}$ 
		& $2.87\times 10^{20}$
		& $8.61\times 10^{15}$ \\
		
		& $\nu = 1\,\text{kHz}$ 
		& Matter code 
		& $<10^{-3}$ 
		& $4.75\times 10^{12}$
		& $1.42\times 10^{8}$ \\
		
		& 
		& Minkowski code 
		& $<10^{-3}$ 
		& $2.89\times 10^{12}$
		& $8.66\times 10^{7}$\\
		\hline
	\end{tabular}
	
	\label{tab_ccsne_damp_DT}
	
\end{table}

\begin{figure}
	\centering
	\includegraphics[scale=0.5]{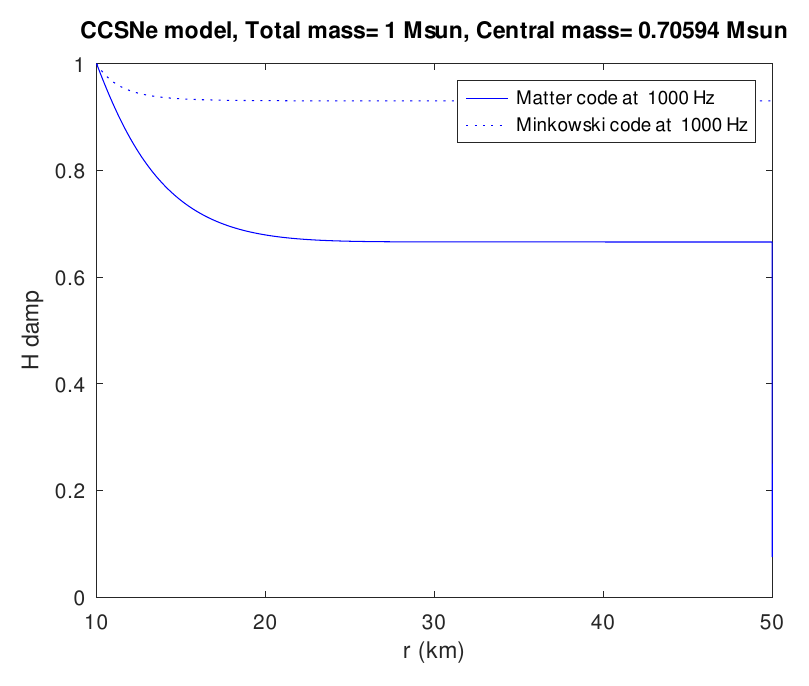}
	\includegraphics[scale=0.5]{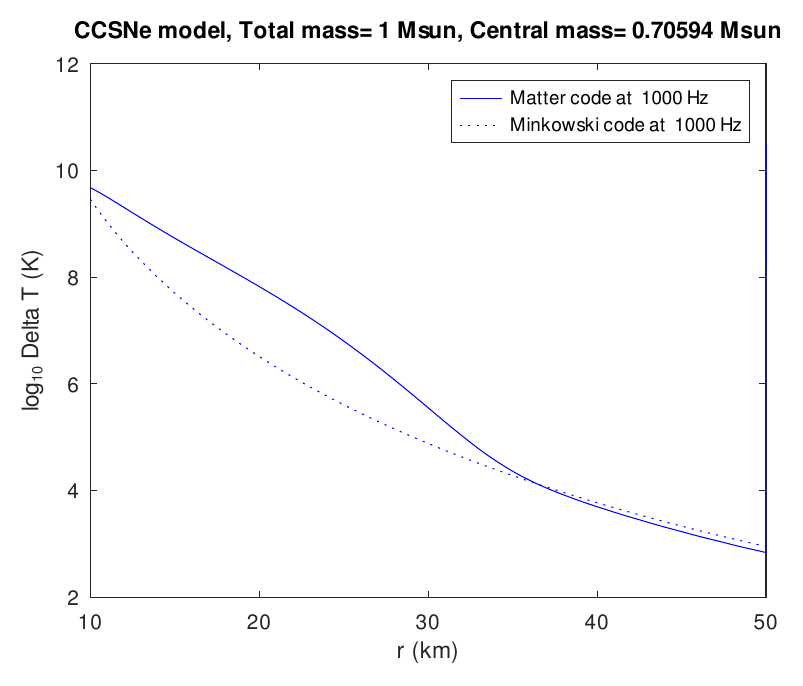}
	\caption{H damp (left) and $\log_{10}(\Delta T)$ (K) (right) plotted against distance (km) for the CCSN model.}
	\label{example_ccsne}
\end{figure}

We evaluate Eqs.~\eqref{e-Ho} and \eqref{e-DeltaT} using the parameter values in Table~\ref{tab:ccsne_parameters} and varying the frequency over $100-1000~\textrm{kHz}$ and the viscosity over $10^{23}$--$10^{25}~\mathrm{kg\,m^{-1}\,s^{-1}}$. For these computations, the total mass is $1M_\odot$, the core mass is $0.71M_\odot$ and the core radius is $10$km. The GW damping factor is reported at the outer boundary and the temperature rise at the inner boundary. The temperature rise depends on $\Delta E_{GW}/C$, and we report the maximum value (when $\Delta E_{GW}=10^{39}\mathrm{J},C=6\mathrm{J\,kg^{-1}\,K^{-1}}$) as well as the minimum value (when $\Delta E_{GW}=10^{36}\mathrm{J},C=200\mathrm{J\,kg^{-1}\,K^{-1}}$). The results are summarized in Table~\ref{tab_ccsne_damp_DT}. In all cases, it is seen that the damping of GWs is almost complete, so we have also found, for various frequencies, a critical value of the shear viscosity, $\eta_c$, such that $10\eta_c$ gives $H_o/H_i<0.1$ and $\eta_c/10$ gives $H_o/H_i>0.9$. The results are:
\begin{equation}
\begin{tabular}{llllll}
\mbox{Frequency (Hz):}& $100$ & $200$ & $500$ &$1000$\\
$\eta_c \mathrm{(kg\,m^{-1}\,s^{-1}}):$  &$10^{14}$&$10^{16}$&$2\times 10^{19}$&$10^{22}$
\end{tabular}\,.
\end{equation}
As an example, Fig.~\ref{example_ccsne} uses $\eta=10^{22}\mathrm{kg\,m^{-1}\,s^{-1}}$ and plots the damping factor and $\log_{10}$ of the temperature rise   against radial distance.

The results obtained have several implications
\begin{itemize}
\item It is important to use the matter code, since results obtained using the Minkowski code always underestimate the GW damping and heating effects.
\item In all models, the GW damping was almost complete. This implies that the GW signal from a CCSNe, even if nearby, would not be observable except possibly in a frequency band greater than $1$kHz.
\item The predicted temperature increases need to be compared to the ambient temperature in the mantle of a CCSNe, which is about $10^{12}$K~\cite{Boccioli:2024abp}, and this value is exceeded in the lower frequency band ($100$Hz). However, the temperature increase occurs deep within the star and is not directly observable. It may be that it could lead to observable effects, but the required modeling is beyond the scope of this paper.
\end{itemize}

%%%%%%%%%%%%%%%%%%%%%%%%%%%%%%%%%%%%%%%%%%%%%%%
\subsection{Binary neutron star merger}
\label{s-bns}

From the first detection of GWs from the binary neutron star merger GW170817 to subsequent events, BNS mergers have been recognized as major GW sources. The parameter estimates presented in Table~\ref{tab:bns_parameters} correspond to the post-merger stage of a BNS system, based on numerical and observational results reported in~\cite{abbott2017gw170817,abbott2017multi,abbott2020gw190425,abbott2017gravitational,hammond2025investigating,fonseca2021refined,antoniadis2013massive,fontbute2025gravitational}. Moreover, these estimates are consistent with those inferred from GW170817~\cite{abbott2019properties,abbott2016observation,abbott2017gw170817,abbott2017gravitational,abbott2020gw190425}. 

Considering both long-lived and short-lived remnants, the core mass lies in the range \(2.4\text{--}2.8~M_\odot\), with a corresponding radius of \(10\text{--}12~\mathrm{km}\). Including the surrounding accretion disk, the total bound mass is \(2.7\text{--}3.1~M_\odot\), with the disk contributing between \(5\times10^{-4}\) and \(0.3~M_\odot\). The density of the shell is modeled as
\begin{equation}
    \rho(r) = \rho_0 \, e^{-r/r_0},
\end{equation}
where \(\rho_0 = 10^{17}~\mathrm{kg\,m^{-3}}\) and \(r_0 = 12~\mathrm{km}\).
\begin{table}[t]
\centering
\caption{Parameter values for the post-merger remnant of a BNS system.}
\label{tab:bns_parameters}
\begin{tabular}{lc}
\hline
\textbf{Quantity} & \textbf{Typical Range or Value} \\
\hline
Core mass & $2.4$--$2.8~M_\odot$ \\
Core radius & $10$--$12~\mathrm{km}$ \\
Total bound mass (core + disk) & $2.7$--$3.1~M_\odot$ \\
Disk mass & $5\times10^{-4}$--$0.3~M_\odot$ \\
Shear viscosity $\eta$ & $10^{24}$--$10^{30}~\mathrm{kg\,m^{-1}\,s^{-1}}$ \\
Specific heat $C$ & $5.84~\mathrm{J\,kg^{-1}\,K^{-1}}$ \\
GW frequency & $1$--$2~\mathrm{kHz}$\\
GW energy release $\Delta E_{\rm GW}$ & $10^{45}~\mathrm{J}$ \\
Density profile & $\rho(r) = \rho_0 e^{-r/r_0}$ \\
\hline
\end{tabular}
\end{table}

\begin{figure}[h]
\centering
\includegraphics[scale=0.5]{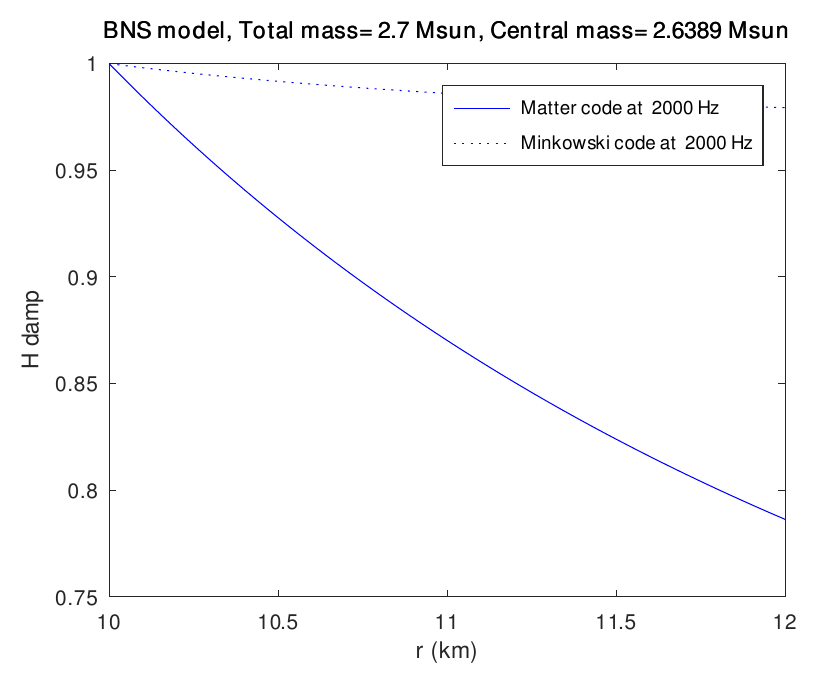}
\includegraphics[scale=0.5]{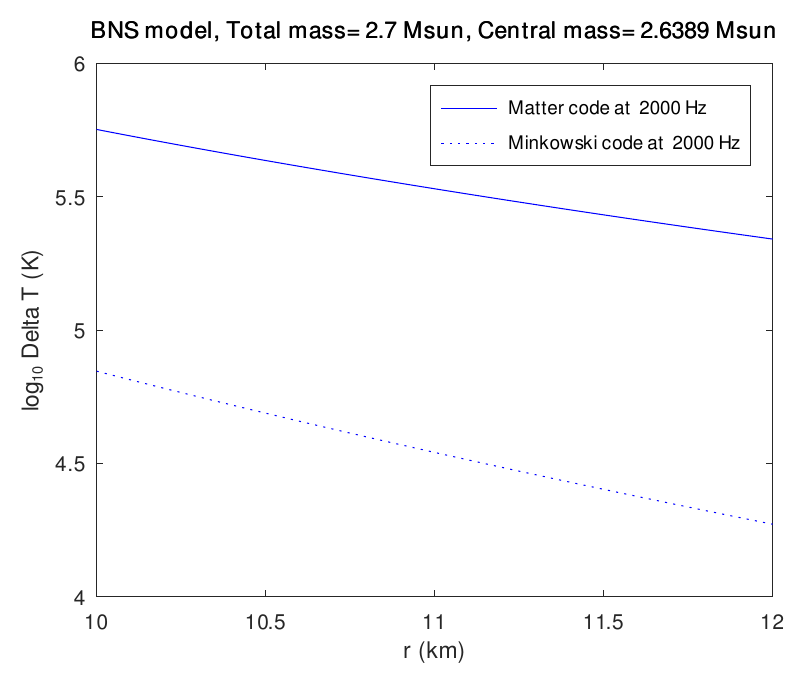}
\caption{H damp (left) and $\log_{10}(\Delta T)$ (K) (right) plotted against distance (km) for the BNS post-merger model.}
\label{example_BNS}
\end{figure}

We evaluate Eqs.~\eqref{e-Ho} and \eqref{e-DeltaT} using the parameters values in Table~\ref{tab:bns_parameters} and varying the frequency over $1-2~\textrm{kHz}$ and the viscosity over $10^{24}$--$10^{30}~\mathrm{kg\,m^{-1}\,s^{-1}}$. For these computations, the total mass is $2.7M_\odot$, the core mass is $2.64M_\odot$ and the core radius is $10$km. The GW damping factor is reported at the outer boundary and the maximum temperature rise at the inner boundary. The results are summarized in Table~\ref{tab_BNS_damp_DT}. The general trend is clear: general spherically symmetric background with high values of viscosity and the frequency exhibit stronger damping and correspondingly larger heating than the Minkowski background. As an example, for $\eta=10^{24}\mathrm{kg\,m^{-1}\,s^{-1}}$ and $\nu=2~\textrm{kHz}$, the damping factor at the outer boundary and the temperature rise at the inner boundary are shown in Fig.~\ref{example_BNS}.

\begin{table}[h]
\caption{The table showing the outer-boundary damping and maximum inner-boundary temperature rise for the BNS models over $r = 10$--$12~\textrm{km}$.}
\label{tab_BNS_damp_DT}
\centering
\renewcommand{\arraystretch}{1.4}
\begin{tabular}{c c l c c}

\hline
$\eta~(\mathrm{kg\,m^{-1}\,s^{-1}})$ & Frequency & Model & $H_{\textrm{damp}}$ (outer) & $\Delta T$ (inner) \\
\hline

% ====================== eta = 10^24 ======================
\multirow{4}{*}{$10^{24}$}

& $\nu = 1\,\text{kHz}$ 
  & Matter code 
  & $<10^{-3}$ 
  & $1.2319\times 10^{8}$ \\

& 
  & Minkowski code 
  & $<10^{-2}$ 
  & $1.6259\times 10^{7}$ \\

& $\nu = 2\,\text{kHz}$ 
  & Matter code 
  & $0.78632$ 
  & $5.6650\times 10^{5}$ \\

& 
  & Minkowski code 
  & $0.97938$ 
  & $7.0194\times 10^{4}$ \\
\hline

% ====================== eta = 10^26 ======================
\multirow{4}{*}{$10^{26}$}

& $\nu = 1\,\text{kHz}$ 
  & Matter code 
  & $<10^{-3}$ 
  & $1.24165\times 10^{10}$ \\

& 
  & Minkowski code 
  & $<10^{-3}$ 
  & $1.73900\times 10^{9}$ \\

& $\nu = 2\,\text{kHz}$ 
  & Matter code 
  & $<10^{-3}$ 
  & $5.4387\times 10^{7}$ \\

& 
  & Minkowski code 
  & $0.12466$ 
  & $6.8501\times 10^{6}$ \\
\hline

% ====================== eta = 10^30 ======================
\multirow{4}{*}{$10^{30}$}

& $\nu = 1\,\text{kHz}$ 
  & Matter code 
  & $<10^{-3}$ 
  & $1.2388\times 10^{14}$ \\

& 
  & Minkowski code 
  & $<10^{-3}$ 
  & $1.6904\times 10^{13}$ \\

& $\nu = 2\,\text{kHz}$ 
  & Matter code 
  & $<10^{-3}$ 
  & $5.5976\times 10^{11}$ \\

& 
  & Minkowski code 
  & $<10^{-3}$ 
  & $7.1121\times 10^{10}$ \\
\hline
\end{tabular}

\end{table}
Overall, Table~\ref{tab_BNS_damp_DT} demonstrates that higher viscosity and lower GW frequency enhance the damping, thereby reducing the fraction of the wave that escapes the remnant system. This implies a reduced detectability for high-viscosity environments, due to strong GW–matter interactions.

Furthermore, it has been reported in the literature that the ambient thermal state of a post-merger BNS system is expected to rise to 
$10-100~\textrm{MeV}$ ($\approx 10^{11}$--$10^{12}$K), primarily due to shock heating~\cite{raithel2021realistic,baiotti2017binary,paschalidis2017rotating}. Remarkably, several cases in Table~\ref{tab_BNS_damp_DT} show that GW-induced viscous heating alone can reach comparable temperatures, implying that dissipation of the post-merger GW signal may contribute non-negligibly to the thermal budget of the remnant.

\subsection{Accretion at a binary black hole merger}
It is expected that most black hole mergers occur in an environment devoid of any significant amount of matter, but it will, on occasion, happen that a merger takes place in an environment where other objects provide a source of matter.
The effect of an accretion disk at a binary black hole merger was considered in~\cite{bishop2024heating} and assumed a Minkowski background. Here, as before, we model the effect based on the parameters of GW150914 \cite{scientific2016tests}
\begin{equation}
\Delta E_{GW}=3M_\odot\,,
f=155 \mbox{Hz}\,,
M_f=62 M_\odot\,,
\end{equation}
where $f$ is the frequency at merger and $M_f$ is the final mass. The energy loss $\Delta E_{GW}$ is for the whole inspiral, and we use instead $2M_\odot$ ($=3.6\times 10^{47}$J) which was the energy radiated away as the frequency increased from 90Hz through peak emission at 132Hz and towards 220Hz as merger gave way to ringdown. In our previous work, the heating effect was estimated using a variable frequency expression, although it was noted that using a fixed frequency of 155Hz (i.e., in the middle of the frequency range) led to very similar results. Here, allowing for a variable frequency would be problematic since the code would need to be run for all the frequency values used; thus we use a fixed frequency of 155Hz.
Parameters of a stationary accretion model, as outlined in ~\cite{shakura1976theory,arai1995accretion,abramowicz2013foundations} are used to model the matter in the accretion disk: at the ISCO (Innermost Stable Circular Orbit), the dynamical viscosity $\eta$ is $3.5\times 10^9$J sec/m${}^3$, the density is $\rho$ as $4$kg/m${}^3$ and the specific heat is $1.43\times 10^4$J/kg/${}^\circ$K. The radius of the ISCO is taken as $r=329$km, being the value for a prograde orbit around a black hole of mass $M_f$ with angular momentum parameter $a=0.67$ (which is the estimated value for the GW150914 remnant).

\begin{table}[h]
\centering
\caption{Temperature increases at $r=329$km (ISCO radius) for models using the Minkowski and Schwarzschild background codes, at the poles and the equator.}
\label{t-BBH}
\begin{tabular}{lll}
\hline
&\textbf{Minkowski code} & \textbf{Schwarzschild code} \\
\hline
Equator ($\theta=\pi/2)$ &$3.47\times 10^8$K&$2.46\times 10^{12}$K\\
Poles ($\theta=0,\,\pi)$ & $1.75\times 10^6$K&$3.77\times 10^6$ K\\
\hline
\end{tabular}
\end{table}

\begin{figure}
\includegraphics[scale=0.5]{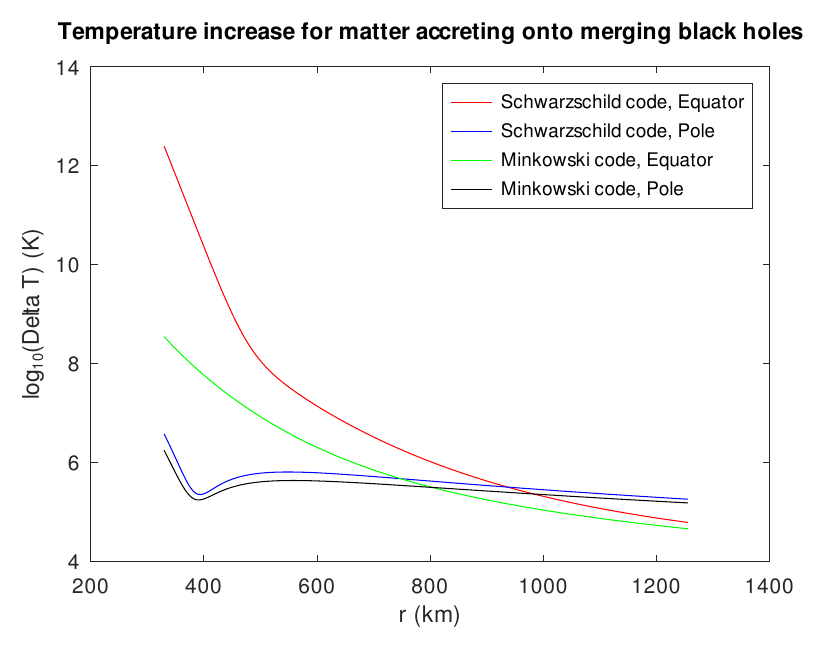}
\caption{$\log_{10}(\Delta T)$ (K) plotted against radial distance (km) for the model of an accretion disk around merging black holes. Plots for both the Schwarzschild background and Minkowski background codes are shown, at both the equator ($\theta=\pi/2$) and the poles ($\theta=0,\pi$)}
\label{f-BBH}
\end{figure}

Since the mass of the accretion disk is insignificant compared to $62M_\odot$, the GW damping and heating effects are computed using the code for a Schwarzschild background; for comparison purposes, results for the Minkowski background are also shown. It was found that GW damping is negligible
\begin{equation}
\frac{H_i-H_o}{H_i} <10^{-14}\,,
\end{equation}
but that the temperature increases can be highly significant, as shown in Fig.~\ref{f-BBH} and Table~\ref{t-BBH}. As noted previously for the Minkowski case~\cite{bishop2024heating}, for small values of $r$ the heating effect is much stronger at the equator (where an accretion disk would be located) than at the poles. The effect is enhanced when using a Schwarzschild background with a temperature increase on the equator at the ISCO of over $10^{12}$K predicted (which is much larger than the ambient temperature of an accretion disk of about $10^6$K). This would be sufficient to generate a gamma-ray burst as may have been observed for GW150914~\cite{Connaughton2016}.

\section{Summary and conclusion}
\label{s-conclusion}
Previous work on the interactions of gravitational waves with matter has been extended to the case of perturbations on a general, static, spherically symmetric background. Generalizing away from a Minkowski background has meant that perturbed quantities no longer have simple forms in terms of elementary functions, and the metric perturbations, and the consequent GW damping and heating effects, are evaluated numerically.

Use of a general spherically symmetric, static background, rather than Minkowski, is more appropriate for many astrophysical problems, and we have, in this paper, applied the theory to core collapse supernovae, the post-merger stage of a binary neutron star merger, as well as to the effect of an accretion disk around merging black holes. In all cases, we found that the GW damping and heating effects were larger than in the case of a Minkowski background, and in some cases substantially larger. Generally, the effects are sufficiently large as to be astrophysically significant for all three scenarios considered. However, if the viscosity is at the bottom of the feasible range, and the frequency at the top of the range, then the astrophysical significance may be only marginal for BNS and perhaps for CCSNe.

It is important to note that the accuracy of the predictions of GW damping and temperature increases are limited by two factors. Firstly, while the background used is more realistic than Minkowski, it is still a crude approximation since the spacetimes being modelled are highly dynamic, and further the GWs cannot properly be treated as small perturbations. Secondly, the temperature increases have taken account of only GW heating, and it can be expected that more realistic models would provide a limit to temperature changes.

An obvious opportunity for future work is to use numerical relativity to calculate fluid shear and the consequent GW damping and heating. For some problems, a Kerr background would be more appropriate than Schwarzschild, but it should be noted that it would be difficult to do so using the Bondi-Sachs formalism since in this case the background metric cannot be expressed in terms of elementary functions~\cite{BishopVenter2006}.

	%%%%%%%%%%%%%%%%%%%%%%%%%%%%%%%%%%%%%%%%%%%%%%%%%
	\begin{acknowledgments}
		This work was supported by the National Research Foundation, South Africa, under grant numbers CPRR240314209194 and RA22111872966.
	\end{acknowledgments}
	\section*{Conflict of interest}
	%%%%%%%%%%%%%%%%%%%%%%%%%%%%%%%%%%%%%%%%%%%%
	The authors declare that they have no conflict of interest.
	%%%%%%%%%%%%%%%%%%%%%%%%%%%%%%%%%%%%%%%%%%%%
	
	\appendix
\section{Schwarzschild interior solution in Bondi-Sachs form}
		\label{a-SI}
The total mass is $M_t$ inside a radius $r_t$ with vacuum for $r>r_t$. The system is at constant density  $\rho$. The background metric and pressure work out to be:
\begin{align}
\rho&=\frac{3M_t}{4\pi r_t^3}\,,\nonumber\\
\beta^{[B]}&=\log\left(\sqrt{\frac{(3\sqrt{1-2M/r_t}-\sqrt{1-2r^2M_t/r_t^3})^2}{4(1-2r^2M_t/r_t^3) }} \right)\,,\nonumber \\
W^{[B]}&=r\left(\exp(-2\beta^{[B]})\left(3\sqrt{1-2M/r_t}-\sqrt{1-2r^2M_t/r_t^3}\right)^2/4-1\right)\,,\nonumber \\
\xi&=\arcsin\left(\sqrt{\frac{2r^2M_t}{r_t^3}}\right)\,,\;
\xi_t=\arcsin\left(\sqrt{\frac{2M_t}{r_t}}\right)\,,\nonumber \\
p&=\rho\frac{\cos(\xi)-\cos(\xi_t)}{3\cos(\xi_t)-\cos(\xi)}\,.
\label{e-PertCon}
\end{align}

	\section{Computer scripts}
	\label{a-MapleOctave}
	
The computer scripts used in this paper are written in plain text format, and are available as Supplemental Material.%~\cite{Sup}. 
The Maple files are in \texttt{Maple.zip} and the Matlab/Octave files are in \texttt{Octave.zip}.
	
\subsection{Maple scripts}
The Maple script \texttt{CCSN.map} calls various other \texttt{.map} scripts and should be run first. Then \texttt{EinsteinBackground.map} creates \texttt{EinsteinBackground.out} which contains expressions used in Eqs.~\eqref{e-sys_back}. The script \texttt{MatlabOutput.map} creates MATLAB/Octave code in the files: \texttt{v1.out} which contains expressions used in Eq.~\eqref{e-dv1}; and \texttt{fE.out}, which has been edited into usable Matlab/Octave code in \texttt{fE\_cleaned.out}, and used to evaluate $f_E$ in Eq.~\eqref{e-fE}. The Maple script \texttt{EinsteinPert.map} should be run on its own and produces \texttt{EinsteinPert.out} which contains expressions used in Eqs~\eqref{e-pertsys},\eqref{e-v0} and \eqref{e-Vang}. The script \texttt{BBH.map} should be run on its own, and creates the file \texttt{BBH\_fE.out} which contains expressions for $f_{E0},f_{E1}$ and $F_{E2}$ used in Eq.~\eqref{e-fE012}.

\subsection{Matlab/Octave scripts}
The files whose names start with \texttt{Driver} call various other \texttt{.m} files and produce figures and other data presented in the paper. Running the file \texttt{DriverBackground.m} compares the background calculated by the code to the interior Schwarzschild solution and produces Fig.~\ref{f-cons}. Running the file \texttt{DriverSchwarzTest.m} compares the perturbed metric fields produced by the Schwarzschild code and the matter code with negligible shell mass, and produces Fig.~\ref{f-Spert}. \texttt{DriverCCSNe.m} and \texttt{DriverBNS.m} produce Figs.~\ref{example_ccsne} and\ref{example_BNS}, respectively, and editing the values of the frequency (\texttt{freq}) and the viscosity (\texttt{eta}) provides the data for Tables~\ref{tab_ccsne_damp_DT} and \ref{tab_BNS_damp_DT}. Running the file \texttt{DriverBBH.m} produces Fig.~\ref{f-BBH} and the data in Table~\ref{t-BBH}.

\begin{comment}	
\section{Navier-Stokes}
\label{a-NS}
Using local proper coordinates, the fluid flow (with $h_\times=0$) is
\begin{equation}
V_i=(-x\dot{h}_+/2,y\dot{h}_+/2,0)
\end{equation}
where $h_+=A\exp(i\omega (t-z/c))=AE$. At linearized level, the Navier-Stokes equation is
\begin{equation}
\rho\frac{\partial V_i}{\partial t}=\rho g_i +\mu\nabla^2 V_i
\end{equation}
and setting $\mu=0$ gives
\begin{equation}
g_i=(-x\ddot{h}_+/2=x\omega^2 AE/2,-y\omega^2AE/2,0)\,.
\end{equation}
Now let $\mu\ne 0$ and $V_i=(-i\omega x BE/2,i\omega y BE/2,0)$. Then Navier-Stokes equation gives
\begin{align}
x&:\; \rho x\omega^2 BE/2=\rho x\omega^2AE/2+\mu x i \omega^3 BE/(2c^2)\nonumber \\
y&:\; -\rho y\omega^2 BE/2=-\rho y\omega^2AE/2-\mu y i \omega^3 BE/(2c^2)\nonumber \\
z&:\; 0=0\,.
\end{align}
In all cases this leads to
\begin{equation}
\rho\omega^2B=\rho\omega^2 A+\frac{i\mu\omega^3B}{c^2}\rightarrow
B=A\left(1-\frac{i\mu\omega}{\rho c^2}\right)^{-1}
\end{equation}
Thus viscosity will significantly modify the flow when 
\begin{equation}
\frac{\mu\omega}{\rho c^2}\gtrapprox 1 \rightarrow 
\frac{2\pi\mu}{\rho\lambda c}\gtrapprox 1 \,.
\end{equation}
For CCSN, $\rho\approx 10^{13}$kg/m${}^3$, $c\approx 3\times 10^8$m/s, $\lambda\approx 3\times 10^5$m and $\mu\approx 10^{23}$kg/m/s, so $2\pi\mu/(\rho\lambda c)\approx 7\times 10^{-4}$.
\end{comment}
\bibliographystyle{plain}
%	\bibliographystyle{spphys}
	% Produces the bibliography via BibTeX.
	\bibliography{tM_1,aeireferences,vis}%,t_1,t,LitRev9a}
	%,aeireferences,t_1,t,LitRev9a}
	%%%%%%%%%%%%%%%%%%%%%%%%%%%%%%%%%%%%%%%%%%%%
\end{document}